\documentclass[twocolumn,showpacs,preprintnumbers,amsmath,amssymb,prb]{revtex4}

\usepackage{graphicx}
\usepackage{dcolumn}
\usepackage{bm}
\usepackage{color}

\begin{document}

\preprint{}

\title{Variety of elastic anomalies in an orbital-active nearly itinerant cobalt vanadate spinel}

\author{Tadataka Watanabe$^1$}
\thanks{tadataka@phys.cst.nihon-u.ac.jp}
\author{Shogo Yamada$^1$}
\author{Rui Koborinai$^2$}
\author{Takuro Katsufuji$^{2,3}$}
\affiliation{$^1$Department of Physics, College of Science and Technology (CST), Nihon University, Tokyo 101-8308, Japan}

\affiliation{$^2$Department of Physics, Waseda University, Tokyo 169-8555, Japan}
\affiliation{$^3$Kagami Memorial Research Institute for Materials Science and Technology, Waseda University, Tokyo 169-0051, Japan}
\date{\today}

\begin{abstract}
We perform ultrasound velocity measurements on a single crystal of nearly-metallic spinel Co$_{1.21}$V$_{1.79}$O$_4$ which exhibits a ferrimagnetic phase transition at $T_C \sim$ 165 K. The experiments reveal a variety of elastic anomalies in not only the paramagnetic phase above $T_C$ but also the ferrimagnetic phase below $T_C$, which should be driven by the nearly-itinerant character of the orbitally-degenerate V 3$d$ electrons. In the paramagnetic phase above $T_C$, the elastic moduli exhibit elastic-mode-dependent unusual temperature variations, suggesting the existence of a dynamic spin-cluster state. Furthermore, above $T_C$, the sensitive magnetic-field response of the elastic moduli suggests that, with the negative magnetoresistance, the magnetic-field-enhanced nearly-itinerant character of the V 3$d$ electrons emerges from the spin-cluster state. This should be triggered by the inter-V-site interactions acting on the orbitally-degenerate 3$d$ electrons. In the ferrimagnetic phase below $T_C$, the elastic moduli exhibit distinct anomalies at $T_1\sim$ 95 K and $T_2\sim$ 50 K, with a sign change of the magnetoresistance at $T_1$ (positive below $T_1$) and an enhancement of the positive magnetoresistance below $T_2$, respectively. These observations below $T_C$ suggest the successive occurrence of an orbital glassy order at $T_1$ and a structural phase transition at $T_2$, where the rather localized character of the V 3$d$ electrons evolves below $T_1$ and is further enhanced below $T_2$.
\end{abstract}


\maketitle

\section{Introduction}

Vanadate spinels $A$V$_2$O$_4$ with divalent $A^{2+}$ ions have attracted considerable attention owing to the interplay between the orbital degree of freedom and geometrical frustration [\cite{Radaelli}]. The trivalent magnetic V$^{3+}$ ions are characterized by double occupancy of the triply-degenerate $t_{2g}$ orbitals ($t_{2g}^2$), and form a sublattice of corner-sharing tetrahedra (pyrochlore lattice). Upon cooling, $A$V$_2$O$_4$, with nonmagnetic $A$ = Zn, Mg, and Cd, undergoes a structural phase transition followed by an antiferromagnetic phase transition [\cite{Ueda,Reehuis,Lee3,Zhang1,Mamiya,Wheeler,Nishiguchi,Onoda,Giovannetti}]. With magnetic $A$ = Mn and Fe, owing to the presence of additional $A^{2+}$-V$^{3+}$ exchange interactions, $A$V$_2$O$_4$ exhibits a more complex structural and magnetic behavior, with successive structural and ferrimagnetic phase transitions [\cite{Adachi,Suzuki,Zhou,Chung,Garlea,Hardy,Nii1,Nii2,Gleason,Katsufuji,MacDougall,Zhang2,Kang,Kawaguchi}].

For $A$V$_2$O$_4$ ($A$ = Zn, Mg, Cd, Mn, and Fe), the structural phase transition is considered to arise from a long-range ordering of the V $t_{2g}$ orbitals, where the lowering of the lattice symmetry should result in the release of frustration (magnetic ordering). The orbital order to explain the structural and magnetic orders for $A$V$_2$O$_4$ is still under debate both theoretically [\cite{Tsunetsugu,Tchernyshyov,Matteo,Marita,Kaur1,Sarkar,Pardo}] and experimentally [\cite{Lee3,Garlea,Wheeler,Suzuki}]; it is considered to be driven by the competition among Jahn-Teller coupling, Kugel-Khomskii exchange interaction [\cite{Kugel}], and relativistic spin-orbit coupling.

CoV$_2$O$_4$ (Co$^{2+}$: $e^4t_2^3$, V$^{3+}$: $t_{2g}^2$) is a unique vanadate spinel with its highest electrical conductivity among the insulating or semiconducting $A$V$_2$O$_4$, which is considered to arise from the nearly-itinerant character of the V 3$d$ electrons [\cite{Kismarahardja,Kiswandhi,Huang,Kaur2,Kismarahardja2,Ma,Koborinai}]. CoV$_2$O$_4$ exhibits a ferrimagnetic order below $T_C\sim$ 150 K, which is similar to $A$V$_2$O$_4$ with magnetic $A$ = Mn ($T_C$ = 57 K) and Fe ($T_C$ = 110 K). However, while $A$V$_2$O$_4$ ($A$ = Zn, Mg, Cd, Mn, and Fe) lowers the cubic lattice symmetry by the structural phase transition, CoV$_2$O$_4$ is unique in that the lowering of the cubic lattice symmetry is absent down to low temperature. Although a very recent neutron scattering study of CoV$_2$O$_4$ discovered a very-weak structural phase transition at $\sim$90 K within the ferrimagnetic phase, this structural transition was characterized as a short-range distortion of the O octahedra, which does not lower the global cubic symmetry of the crystal [\cite{Reig-i-Plessis}]. The very small magnitude of the structural distortion in CoV$_2$O$_4$ is considered to be relevant to the nearly-itinerant electron character.

For CoV$_2$O$_4$, the previous reports of the experimental studies in the single-crystalline and polycrystalline samples suggested the presence of additional transition within the ferrimagnetic phase [\cite{Kiswandhi,Huang,Ma,Koborinai}]. However, the reported transition temperatures, also including the ferrimagnetic transition temperature $T_C$, differ from study to study. This variation in the transition temperature values seems to arise from the off-stoichiometry of the samples. In the crystal growth of the cobalt vanadate spinel, the Co : V ratio of the grown crystal is prone to deviate from the stoichiometric ratio of 1 : 2 to 1+$x$ : 2-$x$ with excess Co $x$. Very recently, the measurements of the magnetization, magnetostriction, neutron scattering, and dielectric constant for the single-crystalline and polycrystalline Co$_{1+x}$V$_{2-x}$O$_4$ with various $x$ extracted the intrinsic magnetic and structural properties irrespective of the off-stoichiometry [\cite{Koborinai}]. It was suggested that, in addition to the ferrimagnetic transition at $T_C$, two other possible transitions occur within the ferrimagnetic phase [\cite{Koborinai}].

Fig. 1(*) depicts the ferrimagnetic transition temperature $T_C$ and two other transition temperatures $T_1$ and $T_2$ in Co$_{1+x}$V$_{2-x}$O$_4$ as functions of off-stoichiometry $x$ (the solid lines), which were suggested from the magnetization and magnetostriction measurements [\cite{Koborinai}]. By increasing the amounts of the off-stoichiometry $x$, $T_C$ increases but $T_1$ and $T_2$ decrease. $T_1$ corresponds to a spin canting temperature, where a collinear-to-noncollinear ferrimagnetic ordering occurs upon cooling [\cite{Koborinai}]. In Fig. 1(*), the structural transition temperature of $\sim$90 K observed in the neutron scattering experiments for the stoichiometric CoV$_2$O$_4$ is also indicated as a circle [\cite{Reig-i-Plessis}], indicating the occurrence of the structural transition at $T_2$.

For $A$V$_2$O$_4$ ($A$ = Mn and Fe), the long-range ordering of the V orbitals is considered to be accompanied by a canting of the V spins, which was observed as the appearance of a noncollinear ferrimagnetic order [\cite{MacDougall,Garlea}]. For Co$_{1+x}$V$_{2-x}$O$_4$, very recent neutron scattering experiments revealed the occurrence of a collinear-to-noncollinear ferrimagnetic ordering (a spin canting) at $T_1$ [Fig. 1(*)] upon cooling to within the ferrimagnetic phase [\cite{Koborinai}]. However, at the spin canting temperature $T_1$, the magnetostriction of Co$_{1+x}$V$_{2-x}$O$_4$ exhibits a change, which is subtler and more gradual than that of $A$V$_2$O$_4$ ($A$ = Mn and Fe) [\cite{Koborinai}]. Furthermore, the dielectric constant of Co$_{1+x}$V$_{2-x}$O$_4$ exhibits a frequency variation within the noncollinear ferrimagnetic phase, indicating slow relaxation dynamics [\cite{Koborinai}]. This dielectric behavior is similar to that observed in the orbital glass state of FeCr$_2$S$_4$ [\cite{Fichtl}]. From the results of the neutron scattering, magnetostriction, and dielectric constant measurements, for the nearly-itinerant Co$_{1+x}$V$_{2-x}$O$_4$, it is suggested that an orbital glassy state exists in the noncollinear ferrimagnetic phase [\cite{Koborinai}].

Herein, we study the interplay of orbital, spin, and lattice degrees of freedom in the nearly-itinerant Co$_{1+x}$V$_{2-x}$O$_4$ by means of ultrasound velocity measurements, which measure elastic moduli of this compound. The elastic modulus of a crystal is a thermodynamic tensor quantity, and thus the ultrasound velocity measurements in all the symmetrically-independent elastic moduli in a crystal can provide the symmetry-resolved thermodynamic information. Furthermore, since the ultrasound velocity can be measured with a high precision of $\sim$ppm, the ultrasound velocity measurements can sensitively probe elastic anomalies driven by phase transition and excitations. For the frustrated spinels, the ultrasound velocity measurements have proven to be a useful tool for studying not only the ground state but also the excited states [\cite{Watanabe1,Watanabe2,Watanabe3,Nii3,Watanabe4,Watanabe5}].

For the vanadate spinels, the ultrasound velocity measurements in MgV$_2$O$_4$ [\cite{Watanabe3}] and MnV$_2$O$_4$ [\cite{Nii3}] were recently reported. These measurements revealed the presence of anomalous elastic behavior in the paramagnetic phase (the magnetically disordered phase) and its absence in the magnetically ordered phase. In contrast, the present study reveals that Co$_{1+x}$V$_{2-x}$O$_4$ exhibits a variety of elastic anomalies in not only the paramagnetic phase but also the magnetically ordered phase. Furthermore, the present study also reveals that the anomalous elastic behavior in the paramagnetic phase of Co$_{1+x}$V$_{2-x}$O$_4$ is uniquely different from that of MgV$_2$O$_4$ and MnV$_2$O$_4$. The present study strongly suggests that the anomalous elastic behavior in Co$_{1+x}$V$_{2-x}$O$_4$ is relevant to the nearly-itinerant character of the orbitally-degenerate V 3$d$ electrons, which causes the orbital glassiness within the magnetically ordered phase.

\section{Experimental}

Single crystals of Co$_{1+x}$V$_{2-x}$O$_4$ ($0\le x\le 0.3$) were prepared by the floating-zone method, where the chemical compositions of the grown single crystals were estimated by the inductively coupled plasma analyses [\cite{Koborinai}]. For the present experiments, we applied a large single crystal of Co$_{1.21}$V$_{1.79}$O$_4$ ($x = 0.21$), where the magnetization, magnetostriction, and neutron scattering measurements suggested the occurrence of magnetic and structural transitions at $T_C\sim$ 165 K, $T_1\sim$ 100 K, and $T_2\sim$ 50 K [the solid lines in Fig. 1(*)] [\cite{Koborinai}]. The ultrasound velocity measurements were performed by a home-built apparatus, where the phase-comparison technique was used with longitudinal and transverse sound waves at a frequency of 30 MHz. The ultrasound waves were generated and detected by LiNbO$_3$ transducers glued on the parallel mirror surfaces of the crystal. We measured the sound velocities in all the symmetrically-independent elastic moduli in the cubic crystal, specifically, compression modulus $C_{11}$, tetragonal shear modulus $\frac{C_{11}-C_{12}}{2}\equiv C_t$, and trigonal shear modulus $C_{44}$. From $C_{11}$ and $C_t$ data, we also obtained the bulk modulus $C_B=\frac{C_{11}+2C_{12}}{3}=C_{11}-\frac{4}{3}C_t$. The respective measurements of $C_{11}$, $C_t$, and $C_{44}$ were performed using longitudinal sound waves with propagation {\bf k}$\parallel$[100] and polarization {\bf u}$\parallel$[100], transverse sound waves with {\bf k}$\parallel$[110] and {\bf u}$\parallel$[1$\bar{1}$0], and transverse sound waves with {\bf k}$\parallel$[110] and {\bf u}$\parallel$[001]. The sound velocities of Co$_{1.21}$V$_{1.79}$O$_4$ measured at room temperature (300~K) are 6740 m/s for $C_{11}$, 2750 m/s for $C_t$, and 3770 m/s for $C_{44}$. In the present study, we also performed the electrical resistivity measurements for the single-crystalline Co$_{1.21}$V$_{1.79}$O$_4$ using the conventional four-probe method.

\begin{figure}[t]
\begin{center}
\includegraphics[scale=0.45]{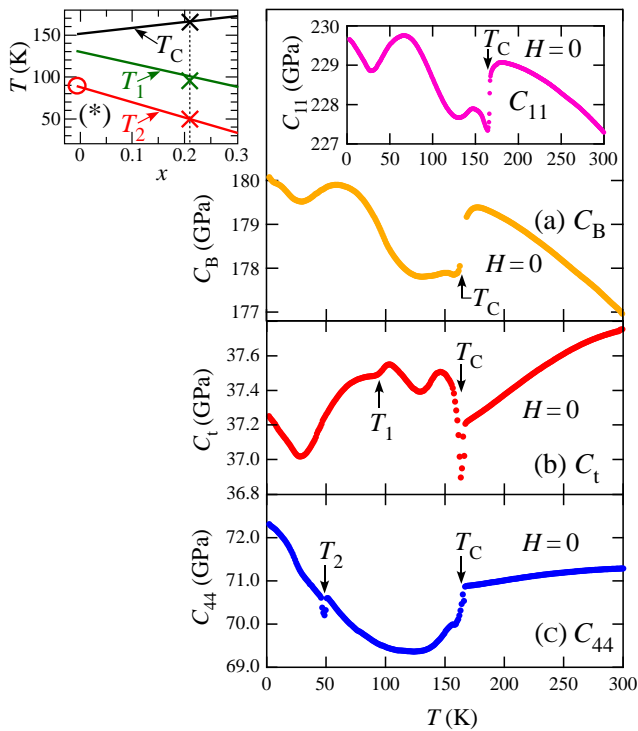}
\caption{\label{fig:fig1} (Color online) (*) Transition temperatures $T_C$, $T_1$, and $T_2$ for Co$_{1+x}$V$_{2-x}$O$_4$ as functions of off-stoichiometry $x$, which were reported in Ref. [\cite{Koborinai}] (the solid lines). The circle in (*) indicates a structural transition temperature for the stoichiometric CoV$_2$O$_4$ reported in Ref. [\cite{Reig-i-Plessis}]. The vertical dashed line in (*) indicates $x$ = 0.21, corresponding to the $x$ value for the single-crystalline sample applied in the present study. The crosses in (*) indicate transition temperatures $T_C$, $T_1$, and $T_2$, which are obtained from the experimental results shown in (a)-(c). (a)-(c) Elastic moduli of Co$_{1.21}$V$_{1.79}$O$_4$ as functions of $T$ with $H$ = 0. (a) $C_B(T)$, (b) $C_t(T)$, and (c) $C_{44}(T)$. The inset in (a) depicts $C_{11}(T)$ of Co$_{1.21}$V$_{1.79}$O$_4$ with $H$ = 0. The arrows in (a)-(c) indicate $T_C$, $T_1$, and $T_2$.}
\end{center}
\end{figure}

\section{Results}
\subsection{Elastic modulus}

Figures 1(a)-1(c) respectively present the temperature ($T$) dependence of the elastic moduli, $C_B(T)$, $C_t(T)$, and $C_{44}(T)$, with zero magnetic field ($H=0$) in Co$_{1.21}$V$_{1.79}$O$_4$. Here, $C_B(T)=C_{11}(T)-\frac{4}{3}C_t(T)$ is obtained from $C_{11}(T)$ [the inset in Fig. 1(a)] and $C_t(T)$ [Fig. 1(b)]. From these experimental results, a variety of elastic anomalies are observed.

First, the elastic moduli shown in Figs. 1(a)-1(c) exhibit anomalies at $\sim$ 165 K for all the elastic moduli, at $\sim$ 95 K for $C_t(T)$, and at $\sim$ 50 K for $C_{44}(T)$. As indicated in Fig. 1(*), these temperatures [the crosses in Fig. 1(*)] respectively agree with $T_C$, $T_1$, and $T_2$ reported in Ref. [\cite{Koborinai}] [the solid lines in Fig. 1(*)]. At the ferrimagnetic transition temperature $T_C$, all the elastic moduli exhibit a discontinuous anomaly, as marked by the arrows in Figs. 1(a)-1(c). Additionally, within the ferrimagnetic phase ($T<T_C$), $C_t(T)$ and $C_{44}(T)$ respectively exhibit a rather gradual anomaly at $T_1\sim$ 95 K and a discontinuous anomaly at $T_2\sim$ 50 K, as marked by the arrows in Figs. 1(b) and 1(c). The gradual anomaly at $T_1\sim$ 95 K in $C_t(T)$ should be driven by the occurrence of the orbital glassy order accompanied by the spin canting, which was suggested in Ref. [\cite{Koborinai}]. The discontinuous anomaly at $T_2$ in $C_{44}(T)$ indicates the occurrence of a phase transition at this temperature. It should be noted that the neutron scattering experiments in the stoichiometric CoV$_2$O$_4$ observed a structural transition at $\sim$ 90 K [the circle in Fig. 1(*)] [\cite{Reig-i-Plessis}]. Since this structural transition temperature agrees with the temperature of $T_2$ for Co$_{1+x}$V$_{2-x}$O$_4$ with $x$ = 0 [the solid line in Fig. 1(*)] [\cite{Koborinai}], the elastic anomaly at $T_2$ in $C_{44}(T)$ for Co$_{1.21}$V$_{1.79}$O$_4$ should be driven by the structural transition, which is identical to that observed in the neutron scattering experiments for the stoichiometric CoV$_2$O$_4$ [\cite{Reig-i-Plessis}]. Comparing the elastic anomalies at $T_C$, $T_1$, and $T_2$, the rather gradual observation of the anomaly at $T_1$ is compatible with the occurrence of the orbital glassy order, which is in contrast to the discontinuous anomalies at $T_C$ and $T_2$ driven by the ferrimagnetic and structural phase transitions, respectively.

In addition to the elastic anomalies at $T_C$, $T_1$, and $T_2$, as shown in Figs. 1(a)-1(c), the elastic moduli exhibit elastic-mode-dependent unusual $T$ variations in both the paramagnetic and ferrimagnetic phases. Upon cooling in the paramagnetic phase ($T>T_C$), $C_B(T)$ exhibits ordinal hardening [\cite{Varshni}], but $C_t(T)$ and $C_{44}(T)$ exhibit anomalous softening. Here, the magnitudes of the softening in $C_t(T)$ and $C_{44}(T)$ are $\Delta C_t/C_t\sim$ 1.5 $\%$ and $\Delta C_{44}/C_{44}\sim$ 0.5 $\%$, respectively. In the ferrimagnetic phase ($T<T_C$), the elastic moduli exhibit nonmonotonic $T$ variations; $C_B(T)$ and $C_t(T)$ exhibit minima at $\sim$ 130 K and $\sim$ 30 K, and $C_{44}(T)$ exhibits a minimum at $\sim$ 130 K.
 
In the present study, we also investigated the magnetic field effect on the elastic properties of Co$_{1.21}$V$_{1.79}$O$_4$. In the ferrimagnetic phase ($T<T_C$), the ultrasound echo signals were too strongly attenuated by the application of a magnetic field to perform the ultrasound velocity measurement, which should arise from the magnetostriction [\cite{Koborinai}]. However, in the paramagnetic phase ($T>T_C$), we were able to detect the ultrasound echo signals independent of whether the magnetic field was applied or not. Thus, in the present study, the ultrasound velocity measurements under magnetic field were performed only in the paramagnetic phase ($T>T_C$).

Figures 2(a)-2(c) respectively depict $C_B(T)$, $C_t(T)$, and $C_{44}(T)$ with magnetic field $H||$[001] in the paramagnetic phase ($T>T_C$) of Co$_{1.21}$V$_{1.79}$O$_4$. Here, $C_B(T)=C_{11}(T)-\frac{4}{3}C_t(T)$ is obtained from $C_{11}(T)$ [the inset in Fig. 2(a)] and $C_t(T)$ [Fig. 2(b)]. In the zero-field ($H=0$) paramagnetic phase, as described above in conjunction with Figs. 1(a)-1(c), $C_t(T)$ and $C_{44}(T)$ exhibit softening with decreasing $T$, while $C_B(T)$ exhibits ordinal hardening [\cite{Varshni}]. Furthermore, in Figs. 2(a)-2(c), $C_{\Gamma}(T)$ exhibits $H$ variation below $\sim$230 K; $C_B(T)$ becomes softer with increasing $H$, but the softening in $C_t(T)$ and $C_{44}(T)$ is relaxed with $H$, as indicated by the solid arrows.

\begin{figure*}[t]
\begin{center}
\includegraphics[scale=0.45]{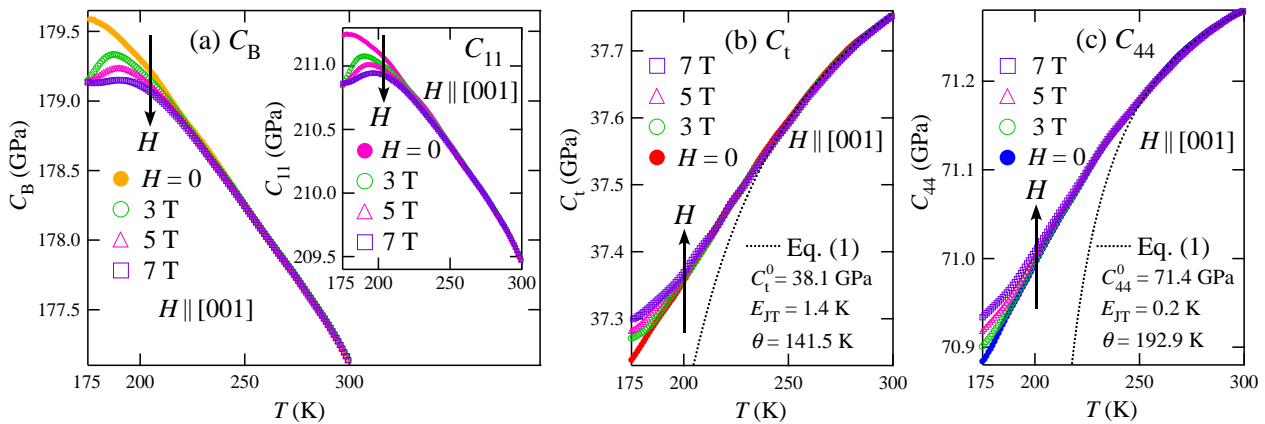}
\caption{\label{fig:fig2} (Color online) $C_{\Gamma}(T)$ of Co$_{1.21}$V$_{1.79}$O$_4$ with $H||$[001] in the paramagnetic phase ($T>T_C$). (a) $C_B(T)$, (b) $C_t(T)$, and (c) $C_{44}(T)$. The inset in (a) depicts $C_{11}(T)$ of Co$_{1.21}$V$_{1.79}$O$_4$ with $H||$[001] in the paramagnetic phase ($T>T_C$). The solid arrows in (a)-(c) are guides to the eye, indicating the variation of $C_{\Gamma}(T)$ with increasing $H$. The dotted curves in (b) and (c) are, respectively, fits of the zero-field experimental $C_t(T)$ and $C_{44}(T)$ to Eq. (1) in 250 K $<T<$ 300 K. The respective values of the fit parameters are also listed in (b) and (c).}
\end{center} 
\end{figure*}

\subsection{Magnetoresistance}

Figure 3(a) depicts the $T$ dependence of the magnetoresistance at 7 T, $MR=\frac{\rho(7~T)-\rho(0)}{\rho(0)}$, for Co$_{1.21}$V$_{1.79}$O$_4$ with the current $I||$[100] and the magnetic field $H||$[001] ($I\perp H$), which is obtained from the $T$ dependence of the electrical resistivities with $H = 0$ ($\rho(0)$) and 7 T ($\rho(7~T)$) depicted in Figs. 3(b) and 3(c). The $T$ dependence of the magnetoresistance shown in Fig. 3(a) agrees with the previous report on the stoichiometric CoV$_2$O$_4$ [\cite{Kismarahardja2}]; upon cooling, Co$_{1+x}$V$_{2-x}$O$_4$ exhibits a large negative magnetoresistance at temperatures around $\sim T_C$, a change to positive magnetoresistance at $\sim T_1$, and a further enhancement of the positive magnetoresistance below $\sim T_2$.

\begin{figure}[t]
\begin{center}
\includegraphics[scale=0.4]{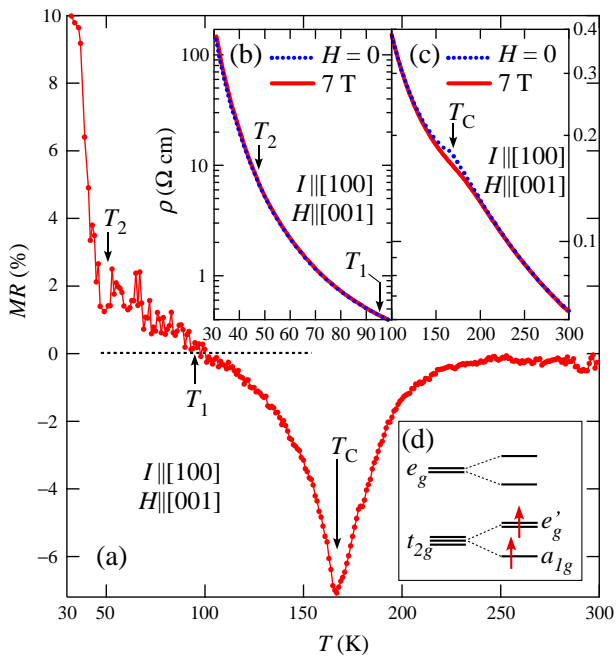}
\caption{\label{fig:fig1} (Color online) (a) Magnetoresistance ($MR$) of Co$_{1.21}$V$_{1.79}$O$_4$ at 7 T for $I||$[100] and $H||$[001] ($I\perp H$) as a function of $T$. (b), (c) Electrical resistivities of Co$_{1.21}$V$_{1.79}$O$_4$ with $H$ = 0 and 7 T ($I||$[100] and $H||$[001]) as functions of $T$: (b) 30 K $<T<$ 100 K and (c) 100 K $<T<$ 300 K. The arrows in (a), (b), and (c) indicate the transition temperatures $T_C$, $T_1$, and $T_2$. The horizontal dashed line in (a) is a guide to the eye indicating $MR=0$. (d) Energy-level scheme of the V 3$d$ orbitals in Co$_{1+x}$V$_{2-x}$O$_4$. The arrows in (d) depict the Hund-rule filling of 2 electrons/V.}
\end{center}
\end{figure}

\section{Discussion}
\subsection{Paramagnetic phase ($T>T_C$)}

We will discuss the origins of the elastic anomalies observed in Co$_{1.21}$V$_{1.79}$O$_4$. First we address the origins of the elastic anomalies in the paramagnetic phase ($T>T_C$) [Fig. 2].

In zero magnetic field ($H$ = 0), $C_t(T)$ and $C_{44}(T)$ soften upon cooling in the paramagnetic phase, while $C_{11}(T)$ exhibits ordinal hardening [\cite{Varshni}]. One probable origin for this softening in $C_t(T)$ and $C_{44}(T)$ is a precursor to the Jahn-Teller (JT) structural transition, which has been observed in MgV$_2$O$_4$ [\cite{Watanabe3}] and MnV$_2$O$_4$ [\cite{Nii3}].

For the JT effect, a JT-active ion and its set of ligands are considered to be a structural unit, where the crystal-field striction mechanism leads to the JT distortion. In the JT magnets, the degenerate ground state is considered to couple strongly and selectively to the elastic modulus $C_{\Gamma}$, which has the same symmetry as the JT distortion. For such a JT-active elastic mode, $T$ dependence of the elastic modulus $C_{\Gamma}(T)$ above the JT transition temperature is explained by assuming the coupling of the ultrasound to the JT-active ions through the crystal-field striction mechanism, and the presence of inter-JT-active-ion interactions. A mean-field expression of $C_{\Gamma}(T)$ in the JT magnets is given as [\cite{Kino, Kataoka, Hazama}]
\begin{equation}
C_{\Gamma}(T) = C_{0,\Gamma} (1-\frac{E_{JT}}{T-\theta}),
\label{eq:JT}
\end{equation}
with $C_{0,\Gamma}$ the elastic constant without the JT effect, $E_{JT}$ the JT coupling energy, and $\theta$ the inter-JT-active-ion interaction.

According to Eq.~(\ref{eq:JT}), $C_{\Gamma}(T)$ of the JT system exhibits a Curie-type ($\sim-1/T$-type) softening at temperatures above the JT transition temperature. For Co$_{1.21}$V$_{1.79}$O$_4$, however, $C_t(T)$ [Fig. 2(b)] and $C_{44}(T)$ [Fig. 2(c)] exhibit the non-Curie-type softening in the paramagnetic phase above $T_C$. As a comparison, fits of the zero-field experimental $C_t(T)$ and $C_{44}(T)$ to Eq.~(\ref{eq:JT}) in 250 K $<T<$ 300 K are presented in Figs. 2(b) and 2(c) as dotted curves, respectively. It is evident in these figures that the experimental $C_t(T)$ and $C_{44}(T)$ both deviate from the dotted curves below $\sim$250 K, which rules out the JT effect as a possible origin for the softening in $C_t(T)$ and $C_{44}(T)$ in the paramagnetic phase above $T_C$. Indeed, for the stoichiometric CoV$_2$O$_4$, the X-ray and neutron diffraction experiments reported the absence of a structural distortion at $T_C$, indicating the absence of a JT structural transition at $T_C$ [\cite{Kismarahardja,Reig-i-Plessis}].

The other possible origin for the softening in $C_t(T)$ and $C_{44}(T)$ is the coupling between the correlated paramagnetic state and the acoustic phonons. It is noted here that the inelastic neutron scattering experiments in the stoichiometric CoV$_2$O$_4$ observed quasielastic magnetic excitations in the paramagnetic phase [\cite{Reig-i-Plessis}]. A similar type of local spin excitations was also observed in $A$V$_2$O$_4$ and $A$Cr$_2$O$_4$ ($A$ = Mg and Zn), which has been characterized as spin-cluster excitations on the V/Cr pyrochlore lattice [\cite{Lee1,Lee2,Tomiyasu1}]. Furthermore, for these compounds, the non-Curie-type softening was observed in $C_{\Gamma}(T)$ in the paramagnetic phase, which is considered to be driven by the coupling of the lattice to the spin-cluster excitations [\cite{Watanabe1,Watanabe3}]. Therefore, the non-Curie-type softening in $C_t(T)$ and $C_{44}(T)$ in the paramagnetic phase of Co$_{1.21}$V$_{1.79}$O$_4$ should stem from the coupling of the lattice to the spin-cluster excitations on the V pyrochlore lattice. Taking into account that the quasielastic magnetic excitations of CoV$_2$O$_4$ and $A$V$_2$O$_4$ ($A$ = Mg and Zn) were both observed centered around a wave vector of $Q\sim 1.3~\AA^{-1}$ [\cite{Reig-i-Plessis,Lee1}], the shape of the spin cluster in these compounds might be identical, which should be confirmed in a future study.

The softening in $C_{\Gamma}(T)$ driven by the spin-cluster excitations is generally explained as the presence of a finite gap for the excitations, which is sensitive to strain [\cite{Watanabe1}]. In the mean-field approximation, $C_{\Gamma}(T)$ in the spin-cluster system is written as [\cite{Watanabe1}]
\begin{equation}
C_{\Gamma}(T)=C_{0,\Gamma}-G_{\Gamma}^2N\frac{\chi_{\Gamma}(T)}{[1-K_{\Gamma}\chi_{\Gamma}(T)]},
\label{eq:SM}
\end{equation}
with $C_{0,\Gamma}$ the background elastic constant, $N$ the density of spin clusters, $G_{\Gamma}=|\partial \Delta/\partial \epsilon_{\Gamma}|$ the coupling constant for a single spin cluster measuring the strain ($\epsilon_{\Gamma}$) dependence of the excitation gap $\Delta$, $K_{\Gamma}$ the inter-spin-cluster interaction, and $\chi_{\Gamma}(T)$ the strain susceptibility of a single spin cluster. From Eq.~(\ref{eq:SM}), when $C_{\Gamma}(T)$ strongly couples to the excited state at $\Delta$, this elastic mode exhibits softening upon cooling roughly down to $T\sim\Delta$, but recovery of the elasticity (hardening) roughly below $T\sim\Delta$; $C_{\Gamma}(T)$ exhibits a minimum roughly at $T\sim\Delta$. According to Eq.~(\ref{eq:SM}), for $C_t(T)$ and $C_{44}(T)$ of Co$_{1.21}$V$_{1.79}$O$_4$ respectively shown in Figs. 2(b) and 2(c), the softening upon cooling down to $T_C$ indicates that this softening arises from the coupling of the lattice to the gapped excitations with $\Delta<T_C$, which is compatible with the observation of the spin-cluster excitations at energies below $\sim$ 4 meV ($< k_BT_C$) in the paramagnetic phase of the stoichiometric CoV$_2$O$_4$ [\cite{Reig-i-Plessis}]. For the experimental results shown in Fig. 2, the future identification of the shape of the spin cluster in Co$_{1+x}$V$_{2-x}$O$_4$ by the inelastic neutron scattering study will enable the quantitative analyses using Eq.~(\ref{eq:SM}).

As observed in Fig. 2, $C_B(T)$, $C_t(T)$, and $C_{44}(T)$ in the paramagnetic phase all exhibit the magnetic-field ($H$) variations. Notably, while the softening in $C_t(T)$ and $C_{44}(T)$ is present already at $H$ = 0 and relaxed with increasing $H$, the softening in $C_B(T)$ is absent at $H$ = 0 but induced by $H$. As described above, the softening in $C_t(T)$ and $C_{44}(T)$ with $H$ = 0 is attributed to the coupling of the lattice to the spin-cluster excitations. Thus the relaxation of this softening by $H$ indicates that the coupling of the lattice to the spin-cluster state is relaxed by $H$. A similar type of this $H$ effect was also observed in MgV$_2$O$_4$, which is also considered to be a result of the relaxation of the spin-cluster-lattice coupling by $H$ [\cite{Watanabe3}]. In contrast, the $H$-induced softening in $C_B(T)$ for Co$_{1.21}$V$_{1.79}$O$_4$ is a unique elastic anomaly, which was not observed in other $A$V$_2$O$_4$ ($A$ = Mg [\cite{Watanabe3}] and Mn [\cite{Nii3}]) compounds. Taking into account that, among the insulating or semiconducting $A$V$_2$O$_4$, Co$_{1+x}$V$_{2-x}$O$_4$ is closest to the itinerant-electron limit [\cite{Kismarahardja,Pardo}], the $H$-induced softening in $C_B(T)$ unique to Co$_{1+x}$V$_{2-x}$O$_4$ should be relevant to the nearly-itinerant character of the V 3$d$ electrons.

As shown in Fig. 3(a), Co$_{1.21}$V$_{1.79}$O$_4$ exhibits the $T$-dependent magnetoresistance, which should arise from the nearly-itinerant V 3$d$ electrons. Comparing Fig. 2(a) with Fig. 3(a) in the paramagnetic phase ($T>T_c$), it is evident that the $H$-induced softening in $C_B(T)$ develops upon cooling below $\sim$230 K in accordance with the development of the negative magnetoresistance. We note here that, for the cubic crystal of Co$_{1+x}$V$_{2-x}$O$_4$, $C_B$ is a symmetry-conserving isotropic elastic mode, while $C_t$ and $C_{44}$ are symmetry-lowering anisotropic elastic modes, and that the application of $H$ causes the development of the softening for $C_B(T)$ and the relaxation of the softening for $C_t(T)$ and $C_{44}(T)$. This elastic-mode-dependent $H$ effect indicates that the magnetoelastic coupling becomes rather isotropic with increasing $H$, which should be driven by the $H$-enhanced delocalization of the strongly-directional V 3$d$ electrons. Thus, for Co$_{1+x}$V$_{2-x}$O$_4$, the $H$-induced softening in $C_B(T)$ and the negative magnetoresistance suggest the presence of the intersite V 3$d$ electron interactions, which causes the $H$-enhanced electron delocalization in the paramagnetic phase.

As is evident from the comparison between Fig. 2 and Fig. 3(a), in the paramagnetic phase of Co$_{1.21}$V$_{1.79}$O$_4$, not only the $H$-induced softening in $C_B(T)$ but also the $H$-induced relaxation of the softening in $C_t(T)$ and $C_{44}(T)$ coincide with the development of the negative magnetoresistance below $\sim$230 K. As already described in conjunction with Figs. 2(b) and 2(c), the softening in $C_t(T)$ and $C_{44}(T)$ with $H$ = 0 in the paramagnetic phase is concluded to be driven by the coupling of the lattice to the spin-cluster state. Thus the $H$-induced relaxation of the softening in $C_t(T)$ and $C_{44}(T)$ suggests the occurrence of the $H$-induced "bond dissolution" in the V spin clusters, which should be driven by the $H$-enhanced delocalization of the V 3$d$ electrons.

In Co$_{1+x}$V$_{2-x}$O$_4$, the O octahedra surrounding the V atoms are trigonally distorted from the regular octahedral shape even in the cubic crystal structure [\cite{Kismarahardja,Kaur2,Reig-i-Plessis}]. Additionally, for the V 3$d$ orbitals, the small trigonal crystal field splits the triply-degenerate $t_{2g}$ orbitals into a localized $a_{1g}$ orbital and delocalized doubly-degenerate $e_g'$ orbitals [Fig. 3(d)]. Thus it is expected that the delocalized $e_g'$ orbitals are responsible for the nearly-itinerant character of the V 3$d$ electrons [\cite{Kaur2}]. The $e_g'$ orbitals are also expected to be responsible for the elastic anomalies in the paramagnetic phase [Fig. 2].

Finally, we note that the softening in $C_{\Gamma}(T)$ driven by the spin-cluster-lattice coupling is observed in not only the orbital-degenerate Co$_{1+x}$V$_{2-x}$O$_4$ and MgV$_2$O$_4$ [\cite{Watanabe3}] but also the orbital-nondegenerate $A$Cr$_2$O$_4$ ($A$ = Mg and Zn) [\cite{Watanabe1}] and ZnFe$_2$O$_4$ [\cite{Watanabe2}], which indicates that the spin-cluster state can universally emerge in the frustrated spinels. It is also noted that this type of elastic softening is relaxed by $H$ in Co$_{1+x}$V$_{2-x}$O$_4$ and MgV$_2$O$_4$, but insensitive to $H$ in $A$Cr$_2$O$_4$ and ZnFe$_2$O$_4$, indicating that the orbital sector is responsible for the $H$ effect on the spin-cluster state. For Co$_{1+x}$V$_{2-x}$O$_4$ and MgV$_2$O$_4$, as described above, it is suggested that the application of $H$ results in the "bond dissolution" in the V spin clusters, which should be driven by the enhancement of the delocalized character for the V 3$d$ electrons. Here, comparing Co$_{1+x}$V$_{2-x}$O$_4$ and MgV$_2$O$_4$, only MgV$_2$O$_4$ exhibits, in addition to the spin-cluster-driven elastic anomalies, the Curie-type softening in $C_{\Gamma}(T)$, which is the JT effect (a precursor to the structural transition) expressed by Eq. (1) [\cite{Watanabe3}]. Taking into account that the JT effect is driven by the coupling of the elastic deformations to the localized magnetic moments, the presence of the JT effect in MgV$_2$O$_4$ but its absence in Co$_{1+x}$V$_{2-x}$O$_4$ indicates that the localized character of the V 3$d$ electrons is stronger in MgV$_2$O$_4$ than Co$_{1+x}$V$_{2-x}$O$_4$, which is compatible with the poorer electrical conductivity of MgV$_2$O$_4$ than Co$_{1+x}$V$_{2-x}$O$_4$ [\cite{Kismarahardja,Pardo}].

\subsection{Ferrimagnetic phase ($T<T_C$)}

In this section, we discuss the origins of the elastic anomalies observed in the ferrimagnetic phase ($T<T_C$) of Co$_{1.21}$V$_{1.79}$O$_4$. As shown in Fig. 1, Co$_{1.21}$V$_{1.79}$O$_4$ exhibits nonmonotonic $T$ dependence of the elastic moduli in the ferrimagnetic phase, which is in contrast to the monotonic $T$ dependence observed in the magnetically ordered phase of other $A$V$_2$O$_4$ ($A$ = Mg [\cite{Watanabe3}] and Mn [\cite{Nii3}]) and $A$Cr$_2$O$_4$ ($A$ = Mg and Zn [\cite{Watanabe1}]). This nonmonotonic elastic behavior in Co$_{1.21}$V$_{1.79}$O$_4$ is expected to be relevant to the nearly-itinerant character of the V 3$d$ electrons.

First, one of the remarkable elastic anomalies shown in Fig. 1 is the elastic moduli minima at $\sim$ 30 K and $\sim$ 130 K. Since there is no additional phase transition at these temperatures, the elastic moduli minima should originate from the coupling of the ultrasound to the magnetic excitations, rather than to the static order. Indeed, in the ferrimagnetic phase, magnon excitations were observed in the inelastic neutron scattering experiments [\cite{Reig-i-Plessis}]. However, in the ferrimagnetic phase of Co$_{1+x}$V$_{2-x}$O$_4$, there remains the possibility for the coexistence of the magnon excitations and the spin-cluster excitations. Thus, at present, we cannot identify which excitation is the origin for the respective elastic moduli minima at $\sim$30 K and $\sim$130 K.

Next, as already mentioned in Sec. III A in conjunction with Fig. 1, the $T$ dependence of the elastic moduli suggests the successive occurrence of the orbital glassy order at $T_1\sim$ 95 K and the structural phase transition at $T_2\sim$ 50 K. As shown in Figs. 1(a)-1(c), the elastic anomaly at $T_1$ is observed only in the tetragonal $C_t(T)$ with $E_g$ symmetry. We note here again that, for Co$_{1+x}$V$_{2-x}$O$_4$, the trigonal crystal field splitting of the V 3$d$ orbitals is present even in the cubic crystal structure [Fig. 3(d)] [\cite{Kismarahardja,Kaur2,Reig-i-Plessis}]. Thus the elastic anomaly at $T_1$ only in $C_t(T)$ suggests that the nearly-itinerant doubly-degenerate $e_g'$ orbitals [Fig. 3(d)] play a dominant role for the occurrence of the orbital glassy order. Further, as shown in Fig. 3(a), the magnetoresistance exhibits a negative-to-positive sign change at $T_1$, which indicates that, in the orbital glassy state, the application of $H$ enhances the localized character of the V 3$d$ electrons. While the negative magnetoresistance in the paramagnetic phase is suggested to be driven by the $H$-induced "bond dissolution" in the V spin cluster, the positive magnetoresistance in the orbital glassy state might be driven by the $H$-induced "bond reinforcement" in the V spin clusters.

For the structural phase transition at $T_2$, from Figs. 1(a)-1(c), the selective observation of the elastic anomaly in the trigonal $C_{44}(T)$ suggests the occurrence of a trigonal lattice distortion at $T_2$. Thus the structural transition at $T_2$ is expected to further enhance the trigonal crystal field splitting of the V 3$d$ orbitals, which is present even in the cubic crystal structure for Co$_{1+x}$V$_{2-x}$O$_4$ [Fig. 3(d)] [\cite{Kismarahardja,Kaur2,Reig-i-Plessis}]. Notably, for Co$_{1+x}$V$_{2-x}$O$_4$, Ref. [\cite{Reig-i-Plessis}] revealed that the structural transition at $T_2$ causes a short-range distortion of the O octahedra, but does not lower the global cubic symmetry of the crystal. Thus the $C_{44}(T)$ anomaly at $T_2$ should be driven by the short-range trigonal lattice distortion. It should be noted that Refs. [\cite{Koborinai}] and [\cite{Reig-i-Plessis}] suggested the existence of the orbital glassy state at temperatures of not only $T_2 < T < T_1$ but also $T < T_2$. Taking into account that the localized electron character at low temperatures of $T < T_2$ was revealed by the observation of the spin gap in the inelastic neutron scattering experiments [\cite{Reig-i-Plessis}], it is implied that the localized electron character in the orbital glassy state is further enhanced below $T_2$, which seems to give rise to the further enhancement of the positive magnetoresistance below $T_2$ [Fig. 3(a)].

\begin{figure}[t]
\begin{center}
\includegraphics[scale=0.45]{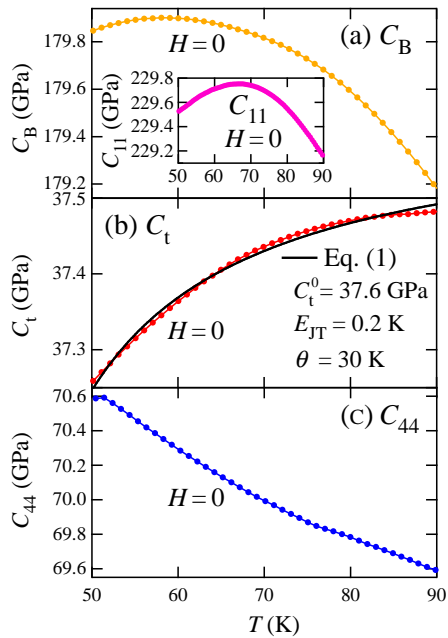}
\caption{\label{fig:fig4} (Color online) $C_{\Gamma}(T)$ of Co$_{1.21}$V$_{1.79}$O$_4$ with $H$ = 0 in $T_2<T<T_1$ [from Figs. 1(a)-1(c)]. (a) $C_B(T)$, (b) $C_t(T)$, and (c) $C_{44}(T)$. The inset in (a) depicts $C_{11}(T)$ of Co$_{1.21}$V$_{1.79}$O$_4$ with $H$ = 0 in $T_2<T<T_1$ [from the inset in Fig. 1(a)]. The solid curve in (b) is a fit of the experimental $C_t(T)$ to Eq.~(1) in 50 K $<T<$ 90 K. The values of the fit parameters are also listed in (b).}
\end{center}
\end{figure}

Finally, we note that, in the orbital glassy phase in the temperature range of $T_2 < T < T_1$, $C_t(T)$ exhibits the Curie-type softening, which is the JT effect expressed by Eq. (1). Figs. 4(a)-4(c) respectively depict $C_B(T)$, $C_t(T)$, and $C_{44}(T)$ in $T_2 < T < T_1$ [from Figs. 1(a)-1(c)]. The Curie-type softening selectively observed in the tetragonal $C_t(T)$ with $E_g$ symmetry should arise from the coupling of the ultrasound to the doubly-degenerate $e_g'$ orbitals [Fig. 3(d)]. Taking into account that the JT effect is driven by the coupling of the elastic deformations to the localized magnetic moments, the presence of the Curie-type softening below $T_1$ is compatible with the existence of the orbital glassy state below $T_1$, where the V 3$d$ electrons should become rather localized. In Fig. 4(b), a fit of the experimental $C_t(T)$ to Eq. (1) is depicted as a solid curve, which reproduces the experimental data very well. Comparing the fit values of the onsite JT energy $E_{JT}$ and the intersite orbital interaction $\theta$ for Co$_{1.21}$V$_{1.79}$O$_4$ [Fig. 4(b)] with those for MgV$_2$O$_4$ [\cite{Watanabe3}] and MnV$_2$O$_4$ [\cite{Nii3}], while $\theta$ for Co$_{1.21}$V$_{1.79}$O$_4$ (30 K) is comparable to those for MgV$_2$O$_4$ (15 K) and MnV$_2$O$_4$ (5 K), $E_{JT}$ for Co$_{1.21}$V$_{1.79}$O$_4$ (0.2 K) is much smaller than those for MgV$_2$O$_4$ (10 K) and MnV$_2$O$_4$ (22 K). Such a small value of $E_{JT}$ for Co$_{1.21}$V$_{1.79}$O$_4$ seems to be compatible with the rather delocalized character of the V 3$d$ electrons compared to MgV$_2$O$_4$ and MnV$_2$O$_4$. Below $T_2$, $C_{\Gamma}(T)$ of Co$_{1.21}$V$_{1.79}$O$_4$ exhibits the absence of the Curie-type softening, and instead exhibits an elastic moduli minimum at $\sim$30 K arising from the magnetic excitations [Figs. 1(a)-1(c)]. For Co$_{1+x}$V$_{2-x}$O$_4$, the JT fluctuations might be suppressed below $T_2$ by the further stabilization of the orbital glassy state, which is accompanied by the lattice distortion.

Although the observation of the Curie-type softening in $C_t(T)$ in $T_2 < T < T_1$ [Fig. 4(b)] indicates the presence of the JT interaction, the $e_g'$ electron [Fig. 3(d)] is also expected to experience the onsite spin-orbit interaction. The very small fit value of $E_{JT}$ for Co$_{1.21}$V$_{1.79}$O$_4$ (0.2 K) [Fig. 4(b)] might reflect, in addition to the rather delocalized electron character, the competition/coexistence of the onsite JT and spin-orbit interactions. Additionally, the disappearance of the Curie-type softening in $C_t(T)$ below $T_2$ [Fig. 1(b)] might be a result of the enhanced contribution of the spin-orbit interaction below $T_2$. For Co$_{1+x}$V$_{2-x}$O$_4$, the intricate interplay of the JT, spin-orbit, and intersite spin-orbital interactions is expected to play an important role for the emergence of the nearly-itinerant electron character and the orbital glassy state, which remains to be understood.

\section{Summary}

Ultrasound velocity measurements of Co$_{1.21}$V$_{1.79}$O$_4$ revealed a variety of the elastic anomalies in both the paramagnetic phase ($T>T_C$) and the ferrimagnetic phase ($T<T_C$). In the paramagnetic phase above $T_C$, the present study revealed the elastic-mode-dependent unusual temperature variations of the elastic moduli, suggesting the existence of the dynamic spin-cluster state. Furthermore, above $T_C$, the present study revealed the sensitive magnetic-field response of the elastic moduli, suggesting that, with the negative magnetoresistance, the magnetic-field-enhanced nearly-itinerant character of the V 3$d$ electrons emerges from the spin-cluster state, which should be triggered by the inter-V-site interactions acting on the orbitally-degenerate 3$d$ electrons. In the ferrimagnetic phase below $T_C$, the elastic anomalies at $T_1\sim$ 95 K and $T_2\sim$ 50 K were found to coincide, respectively, with the sign change of the magnetoresistance at $T_1$ (positive below $T_1$) and the enhancement of the positive magnetoresistance below $T_2$. These observations below $T_C$ suggest the successive occurrence of the orbital glassy order at $T_1$ and the structural phase transition at $T_2$, where the rather localized character of the V 3$d$ electrons evolves below $T_1$ and is further enhanced below $T_2$. Further experimental and theoretical studies are indispensable if the spin and orbital states of the nearly-metallic Co$_{1+x}$V$_{2-x}$O$_4$ in both the magnetically disordered and ordered phases are to be understood.

\section{Acknowledgments}

This work was partly supported by a Grant-in-Aid for Scientific Research (C) (Grant No. 17K05520) from MEXT of Japan, and by Nihon University College of Science and Technology Grants-in-Aid for Project Research.

\end{document}